\documentclass[prd,onecolumn,nofootinbib,preprintnumbers]{revtex4-2}

\usepackage[T1]{fontenc}
\usepackage[utf8]{inputenc}
\usepackage{lmodern}
\usepackage{amsmath,amssymb,mathtools}
\usepackage{bm}
\usepackage{physics}
\usepackage{siunitx}
\usepackage{graphicx}
\usepackage{xcolor}
\usepackage{booktabs,multirow}
\usepackage{enumitem}
\usepackage[colorlinks=true,linkcolor=blue,citecolor=blue,urlcolor=blue]{hyperref}
\usepackage[capitalize,nameinlink]{cleveref}

\begin{document}
\title{Leptophilic Gauge Interactions in the SMEFT Era}

\author{S. O. Kara}
\email[]{seyitokankara@gmail.com} 
\affiliation{Niğde Ömer Halisdemir University, Bor Vocational School, 51240, Niğde, Türkiye}

\begin{abstract}
We develop a \emph{precision–ready}, model–independent EFT framework that connects 
general leptophilic gauge interactions to their low–energy manifestations in SMEFT and LEFT.
Starting from a broad $U(1)'_{\ell}$ extension in which leptons carry family–dependent charges 
$(q_e, q_\mu, q_\tau)$ while quarks remain neutral, quantum consistency is ensured through a
minimal set of vectorlike leptons—chiral under $U(1)'_{\ell}$ but vectorlike under the SM—
together with singlet scalars responsible for symmetry breaking and the masses of both $Z_{\ell}$
and the heavy leptons.

In the heavy–mediator limit, we integrate out $Z_{\ell}$ at tree level and derive compact,
analytic SMEFT Wilson coefficients for four–lepton operators and Higgs–current structures,
including hypercharge--$U(1)'_{\ell}$ kinetic mixing proportional to ${\rm Tr}(YQ')$.
Renormalization–group evolution down to the electroweak scale and matching onto LEFT
produce closed–form expressions directly applicable to 
$e^{+}e^{-}\!\to\!\ell^{+}\ell^{-}$,
neutrino trident production,
$\nu e$ scattering,
parity–violating M\o ller observables,
and muon–decay parameters.

A single parameter combination,
\[
\Lambda_{\rm eff}
    = \frac{M_{Z_\ell}}{g_\ell\sqrt{|q\,q'|}},
\]
governs all leading EFT signatures, offering a unified way to map UV charge assignments
onto precision observables.  
We delineate the validity domain of this EFT and outline its use as a ``plug–and–play''
interface for global fits, collider recasts, and future–collider projections.

\end{abstract}


\maketitle

\section{Introduction}\label{sec:intro}

The observation of neutrino oscillations~\cite{PDG2020} established that the individual lepton flavour numbers 
$L_e$, $L_\mu$ and $L_\tau$ are not conserved, providing one of the clearest indications of physics beyond the Standard Model (SM).  
Among the theoretically economical extensions motivated by this fact, an additional Abelian gauge factor acting on leptons stands out.  
In this work we consider the most general leptophilic $U(1)'_\ell$, under which the three lepton families carry arbitrary charges 
$(q_e,q_\mu,q_\tau)$ while quarks remain neutral.  
Special cases include the flavour-universal assignment $q_e=q_\mu=q_\tau$ and the anomaly-free $L_\mu-L_\tau$ choice $(0,+1,-1)$~\cite{Heeck:2011wj}.  
The corresponding gauge boson, which we denote by $Z_\ell$, couples to leptons at tree level and provides a clean portal to 
precision electroweak, neutrino, and charged-lepton observables.

Such gauge extensions are theoretically well-motivated~\cite{PDG2024} and phenomenologically testable across a wide energy range.  
However, a fully consistent construction requires anomaly cancellation, a viable symmetry-breaking sector, and control over possible 
gauge--kinetic mixing.  
Once these ingredients are specified, the low-energy effects of the heavy mediator may be described in a model-independent way using 
the Standard Model Effective Field Theory (SMEFT) and its low-energy limit, the LEFT.  
Leptophilic interactions are particularly attractive in this context: they produce a restricted set of purely leptonic operators, 
allowing analytic matching and transparent parametric control.

\medskip

In this paper we develop a complete and UV-agnostic EFT bridge between leptophilic gauge interactions and their manifestations in 
precision observables.  
Our analysis proceeds in four steps:  
(i)~we formulate an anomaly-free leptophilic $U(1)'_\ell$ with a minimal set of vectorlike leptons;  
(ii)~we integrate out the heavy gauge boson $Z_\ell$ and derive compact analytic SMEFT Wilson coefficients for four-lepton and 
Higgs--current operators, including kinetic mixing proportional to $\mathrm{Tr}(YQ')$;  
(iii)~we evolve these coefficients to the electroweak scale and match them onto LEFT; and  
(iv)~we identify the resulting patterns in key high- and low-energy probes such as 
$e^+e^-\!\to\ell^+\ell^-$, neutrino trident production, $\nu e$ scattering, parity-violating Møller observables, and muon decay.

A central outcome of our analysis is the emergence of a single effective interaction scale,
\begin{equation}
    \Lambda_{\rm eff}
    = \frac{M_{Z_\ell}}{g_\ell\,\sqrt{|q\,q'|}},
\end{equation}
which controls the size of all induced SMEFT and LEFT operators.  
This scale provides a unified interpretation framework for current and future precision data---including those from LEP, MOLLER, 
neutrino facilities, and upcoming lepton colliders—and enables direct comparison of broad classes of UV completions.

\paragraph*{Organization.}  
\Cref{sec:model} introduces the gauge structure, vectorlike spectrum, and anomaly constraints.  
\Cref{sec:matching} presents the tree-level SMEFT matching, 
\cref{sec:RGE} the RG evolution and matching onto LEFT, 
and \cref{sec:pheno} the resulting phenomenology.  
\Cref{sec:bench} summarizes benchmark scenarios, while \cref{sec:validity} discusses 
the validity of the EFT and theoretical consistency.  
We conclude in \cref{sec:concl}.  
Technical derivations are collected in the Appendices.

\section{General leptophilic model and anomaly cancellation}
\label{sec:model}

\subsection{Field content and charges}
We extend the SM gauge group as
\[
SU(3)_C \times SU(2)_L \times U(1)_Y 
\;\longrightarrow\;
SU(3)_C \times SU(2)_L \times U(1)_Y \times U(1)'_\ell .
\]
Leptons carry family–dependent charges 
\((q_e,q_\mu,q_\tau)\) under \(U(1)'_\ell\),
while quarks remain neutral, realizing a genuinely ``leptophilic’’ gauge interaction
\cite{Langacker:2008yv,Heeck:2014zfa}.
The associated gauge boson is denoted \(Z_\ell\) with coupling \(g_\ell\).
Spontaneous symmetry breaking is achieved through one or more SM–singlet scalar fields 
\(S_i\) with charges \(Q_{S_i}\), whose vacuum expectation values generate
\[
M_{Z_\ell}
  = g_\ell\sqrt{\sum_i Q_{S_i}^2\,v_{S_i}^2},
\]
and induce masses for the heavy vectorlike leptons via Yukawa terms
\(y_S\,S_i\,\overline{L}'_L L'_R\).
This setup represents the minimal UV structure required to produce a well–defined low–energy EFT.

\subsection{Anomaly conditions}

We work in a left-handed Weyl basis so that all chiralities transform consistently.
The vanishing of gauge and mixed anomalies requires
\begin{align}
\mathcal{A}_{[U(1)'_\ell]^3} &= 0, \\
\mathcal{A}_{[SU(2)_L]^2 U(1)'_\ell} &= 0, \\
\mathcal{A}_{[U(1)_Y]^2 U(1)'_\ell} &= 0, \\
\mathcal{A}_{U(1)_Y [U(1)'_\ell]^2} &= 0, \\
\mathcal{A}_{\text{grav}-U(1)'_\ell} &= 0.
\end{align}
These relations yield linear constraints among the $U(1)'_\ell$ charges of SM leptons and the additional vectorlike fields.  
A minimal anomaly-free completion can be achieved by adding one vectorlike generation $(L',E',N')$ that is vectorlike under the SM but chiral under $U(1)'_\ell$, leading to the compact condition
\[
q_e + q_\mu + q_\tau = 0.
\]
as emphasized in 
\cite{Foot:1990mn,Heeck:2011wj}.
Kinetic mixing between the two Abelian factors,
\(-\tfrac{\epsilon}{2} B_{\mu\nu} X^{\mu\nu}\),
is unavoidably induced at one loop \cite{Holdom:1985ag}.
Radiative stability of a small \(\epsilon\) follows if the trace condition 
\(\mathrm{Tr}(Y Q')=0\) holds at the field–theory level, a feature naturally realized in several of our benchmark charge assignments.

\subsection{Masses and interactions}
After symmetry breaking, the charged– and neutral–lepton mass matrices take the schematic form
\[
\mathcal{L}_{\rm mass}
   \supset 
   \overline{E}_L M_E E_R 
 + \overline{N}_L M_N N_R 
 + \text{h.c.},
\]
where \(M_{E,N}\) receive both SM–like and singlet–induced contributions.
Biunitary diagonalization,
\(
E_{L,R}=U^{E}_{L,R}E'_{L,R},
\;
N_{L,R}=U^{N}_{L,R}N'_{L,R},
\)
yields the couplings of \(Z_\ell\) to mass eigenstates:
\begin{align}
\mathcal{L}\supset 
g_\ell Z_{\ell\mu}\Big[
\overline{E}'\,\gamma^\mu (Q^E_L P_L + Q^E_R P_R)\,E'
+
\overline{N}'\,\gamma^\mu (Q^N_L P_L + Q^N_R P_R)\,N'
\Big].
\end{align}
Mixing between SM leptons and the heavy vectorlike states is constrained by electroweak precision tests and is taken to satisfy
\(\sin\theta_{\rm mix}\lesssim10^{-2}\).
Consequently their impact on effective operators enters only at higher order
(see e.g.~\cite{Ellis:2018xal,Bauer:2018onh}).
Explicit charge solutions and mixing formulas are collected in 
\cref{app:anomalies,app:VLL}.

\subsection{Connection to the EFT description}
In the low–energy regime
\(E\ll M_{Z_\ell},M_{L',E',N'}\),
the heavy fields can be integrated out,
leading to contact interactions among SM leptons of the schematic form
\[
\frac{g_\ell^2}{M_{Z_\ell}^2}
(\bar{\ell}\gamma_\mu\ell)\,(\bar{\ell}'\gamma^\mu\ell')\,.
\]
These represent the leading \(d=6\) contributions in the SMEFT expansion and form the starting point for the matching analysis in \cref{sec:matching}.
This section therefore establishes the bridge between the anomaly–free UV theory and its precision–EFT representation.

\section{Tree-level matching onto SMEFT}
\label{sec:matching}

In the regime $p^2 \ll M_{Z_\ell}^2$, the heavy mediator $Z_\ell$ can be integrated out at tree level, leading to a universal and compact set of dimension-six operators.  
The starting point is the leptophilic current
\begin{align}
J'_\mu =
\sum_{\alpha=e,\mu,\tau}
\big[
q^L_\alpha\,\overline{L}_\alpha\gamma_\mu L_\alpha
+ q^e_\alpha\,\overline{e}_\alpha\gamma_\mu e_\alpha
\big],
\end{align}
whose quadratic contraction arises from the equation of motion for $Z_\ell$,
\begin{align}
\mathcal{L}_{\rm eff}^{(6)}
\simeq
-\,\frac{g_\ell^2}{2M_{Z_\ell}^2}\,J'_\mu J'^{\,\mu}.
\label{eq:JJ}
\end{align}
This interaction generates the Warsaw-basis four-lepton structures
$\mathcal{O}_{ll}$, $\mathcal{O}_{le}$, and $\mathcal{O}_{ee}$ 
\cite{Grzadkowski:2010es}, with family-diagonal coefficients
\begin{align}
\frac{[C_{ll}]_{\alpha\alpha\beta\beta}}{\Lambda^2}
&= -\frac{g_\ell^2}{M_{Z_\ell}^2}\,q^L_\alpha q^L_\beta, &
\frac{[C_{le}]_{\alpha\alpha\beta\beta}}{\Lambda^2}
&= -\frac{g_\ell^2}{M_{Z_\ell}^2}\,q^L_\alpha q^e_\beta, &
\frac{[C_{ee}]_{\alpha\alpha\beta\beta}}{\Lambda^2}
&= -\frac{g_\ell^2}{M_{Z_\ell}^2}\,q^e_\alpha q^e_\beta.
\label{eq:SMEFT-4L}
\end{align}

Beyond these current–current operators, 
hypercharge–$U(1)'_\ell$ kinetic mixing 
$-\tfrac{\epsilon}{2} B_{\mu\nu}X^{\mu\nu}$
induces Higgs–current structures whenever 
${\rm Tr}(YQ')\neq0$.  
Integrating out $Z_\ell$ in this case produces
\cite{Holdom:1985ag,delAguila:2010mx}
\begin{align}
\mathcal{O}^{(1)}_{H\ell,\,\alpha}
=(H^\dagger i\overleftrightarrow{D}_\mu H)(\bar L_\alpha\gamma^\mu L_\alpha), 
\qquad
\mathcal{O}_{He,\,\alpha}
=(H^\dagger i\overleftrightarrow{D}_\mu H)(\bar e_\alpha\gamma^\mu e_\alpha),
\end{align}
with leading coefficients
\begin{align}
\frac{[C_{H\ell}^{(1)}]_\alpha}{\Lambda^2}
\simeq 
-\,\epsilon\,\frac{g_Y g_\ell}{M_{Z_\ell}^2}\,Y_H\,q^L_\alpha,
\qquad
\frac{[C_{He}]_\alpha}{\Lambda^2}
\simeq 
-\,\epsilon\,\frac{g_Y g_\ell}{M_{Z_\ell}^2}\,Y_H\,q^e_\alpha,
\qquad (Y_H=+1).
\end{align}

Taken together, these operators encapsulate the complete low-energy imprint of a heavy leptophilic gauge boson.  
The four-lepton terms dominate high-energy processes such as 
$e^+e^-\!\to\!\ell^+\ell^-$ and neutrino trident production, 
while the Higgs–current operators govern electroweak precision observables, parity-violating Møller scattering, and the small induced couplings of $Z_\ell$ to quarks through $Z$–$Z_\ell$ mixing 
\cite{Efrati:2015eaa,Falkowski:2019xoe}.  
These expressions therefore form the backbone for the RGE evolution and LEFT matching discussed in \Cref{sec:RGE}, and constitute the universal EFT dictionary for precision probes of leptophilic interactions.

\section{Running and matching to LEFT}
\label{sec:RGE}

The tree–level SMEFT coefficients obtained at the matching scale $\mu = M_{Z_\ell}$ 
do not directly correspond to the low–energy observables measured in neutrino, 
M{\o}ller, or charged–lepton scattering experiments. 
Between $M_{Z_\ell}$ and the electroweak scale $m_Z$, the coefficients evolve according to 
the full one–loop anomalous–dimension matrices of SMEFT 
\cite{Jenkins:2013wua,Jenkins:2013zja,Alonso:2013hga}. 
This renormalization–group evolution induces electroweak corrections, 
including mild flavor mixing between the four–lepton operators and the Higgs–current 
structures generated by kinetic mixing. 
The running preserves the simple charge–factor dependence of the EFT, 
so that each Wilson coefficient remains proportional to combinations such as 
$q^L_\alpha q^L_\beta$ or $q^L_\alpha q^e_\beta$ throughout the evolution.

At the electroweak scale $\mu = m_Z$, the heavy SM fields $W$, $Z$, and $H$ are 
integrated out.  
Using the standard LEFT basis of Refs.~\cite{Aebischer:2015fzz,Aebischer:2017ugx}, 
the SMEFT operators at $m_Z$ match onto a set of purely leptonic 
and semileptonic LEFT operators.  
For leptophilic gauge extensions, the dominant structures are the 
vector–vector interactions,
\begin{align}
\mathcal{O}^{V,LL}_{\nu_\alpha \ell_\beta}
&=(\bar\nu_\alpha\gamma_\mu P_L\nu_\alpha)
  (\bar\ell_\beta\gamma^\mu P_L\ell_\beta), 
&
\mathcal{O}^{V,LR}_{\nu_\alpha \ell_\beta}
&=(\bar\nu_\alpha\gamma_\mu P_L\nu_\alpha)
  (\bar\ell_\beta\gamma^\mu P_R\ell_\beta),
\end{align}
whose Wilson coefficients inherit the same parametric suppression 
\[
C \;\sim\; \frac{g_\ell^2}{M_{Z_\ell}^2}\,
(q^L_\alpha q^L_\beta,\, q^L_\alpha q^e_\beta),
\]
up to small loop corrections from the SMEFT running.
These operators provide a unified EFT description of several key probes:  
neutrino–electron scattering, neutrino trident production, parity–violating 
M{\o}ller asymmetries, and the extraction of $G_F$ from muon decay.  
Because they interfere directly with SM weak interactions, these observables offer 
some of the most sensitive tests of leptophilic gauge sectors 
in the region well below the $Z_\ell$ pole.

The complete analytic expressions for the LEFT coefficients, along with the 
$G_F$–scheme subtleties and subleading kinetic–mixing effects, are collected in 
\cref{app:LEFT}.  
This scale–separated treatment ensures that the final low–energy EFT retains the 
model–independent structure emphasized throughout this work, enabling direct 
application to precision data and future global analyses.

\section{Phenomenology in the EFT limit}
\label{sec:pheno}

Because $\Lambda_{\rm eff}$ controls all Wilson coefficients in the EFT limit, the phenomenology across all channels can be discussed in a unified manner. The EFT obtained after integrating out $Z_\ell$ predicts a set of 
leptonic contact interactions whose leading effects scale as 
$g_\ell^2/M_{Z_\ell}^2$ times charge products.  
Across all observables, the phenomenology is governed by the 
effective interaction scale 
\(
\Lambda_{\rm eff}=M_{Z_\ell}/(g_\ell\sqrt{|q\,q'|}),
\)
which directly maps onto LEP-II bounds and future lepton-collider sensitivities.
Below we summarize the key probes in the EFT regime $E \ll M_{Z_\ell}$.

\subsection{$e^+e^-\to \ell^+\ell^-$ and contact interactions}

In terms of the standard LEP chiral contact structures 
$(\bar e_i\gamma_\mu e_i)(\bar\ell_j\gamma^\mu \ell_j)$ with $i,j=L,R$,
the effective interactions induced by $Z_\ell$ exchange correspond to
\begin{align}
\frac{4\pi}{\Lambda_{LL,\,e\beta}^2}\eta_{LL}^{e\beta} 
&= -\,\frac{g_\ell^2}{M_{Z_\ell}^2}\,q^L_e q^L_\beta, &
\frac{4\pi}{\Lambda_{LR,\,e\beta}^2}\eta_{LR}^{e\beta}
&= -\,\frac{g_\ell^2}{M_{Z_\ell}^2}\,q^L_e q^e_\beta, &
\text{etc.}
\end{align}
Interference with $\gamma/Z$ exchange modifies both the differential 
cross section and the forward–backward asymmetry $A_{\!FB}$ linearly in $s/M_{Z_\ell}^2$.
This mapping allows for a direct reinterpretation of existing LEP-II limits 
\cite{LEP:2003aa,Efrati:2015eaa,Falkowski:2019xoe}.
Future $e^+e^-$ machines (ILC, FCC-ee, CEPC) can extend the reach in 
$\Lambda_{\rm eff}$ by nearly an order of magnitude.

\subsection{Neutrino trident production and $\nu e$ scattering}

For $\nu_\mu N\!\to\!\nu_\mu\,\mu^+\mu^- N$, integrating out $Z_\ell$ yields
\begin{align}
\mathcal{L}_{\text{eff}}\supset
\frac{g_\ell^2}{M_{Z_\ell}^2}
(\bar\nu_\mu\gamma_\mu P_L\nu_\mu)\,
\big[\bar\mu\gamma^\mu (\alpha_L P_L+\alpha_R P_R)\mu\big],
\end{align}
with coefficients 
$\alpha_L \propto q^L_\mu q^L_\mu$ 
and 
$\alpha_R \propto q^L_\mu q^e_\mu$.  
The resulting interference modifies the SM prediction at order 
$\mathcal{O}(g_\ell^2/M_{Z_\ell}^2)$,  
providing a powerful probe of muon-philic scenarios 
\cite{Altmannshofer:2014pba,Altmannshofer:2019zhy}.

Neutrino–electron scattering constrains the combinations 
$(q^L_e,q^e_e)$ through both the four-lepton operators of 
\cref{sec:matching} and the Higgs–current operators 
$\mathcal{O}_{H\ell}^{(1)}$ and $\mathcal{O}_{He}$.
Current bounds are dominated by CHARM-II and TEXONO data,  
complementing the $e^+e^-$ constraints.

\subsection{Parity-violating M{\o}ller scattering and muon decay}

The operators $C_{H\ell}^{(1)}$ and $C_{He}$ shift the weak charges 
and hence modify parity-violating observables in polarized 
electron–electron scattering, such as the E158 result and 
the upcoming MOLLER experiment 
\cite{Anthony:2005pm,Benesch:2014bas}.
The same structures, together with $C_{ll}$ and $C_{le}$,  
induce universal corrections to muon decay and to the 
extraction of $G_F$, scaling as $g_\ell^2 v^2/M_{Z_\ell}^2$.  
These effects are small but become relevant in high-precision fits.

\subsection{Muon $(g-2)$: qualitative expectations}

At tree level no dipole operator is generated.  
Loop-induced SMEFT renormalization mixes the four-lepton and 
Higgs–current operators into $\mathcal{O}_{e\gamma}$, leading to the 
parametric estimate
\begin{align}
\Delta a_\mu \sim 
\frac{m_\mu^2}{16\pi^2}\,
\frac{g_\ell^2}{M_{Z_\ell}^2}\,
(\text{charge factors})\,
\ln\!\frac{M_{Z_\ell}^2}{m_\mu^2},
\end{align}
as discussed in \cref{app:dipole}.
A full computation requires specifying the vectorlike-lepton sector 
and associated thresholds 
\cite{Jegerlehner:2009ry,Endo:2021zal}.  
Within the pure EFT, $\Delta a_\mu$ remains loop- and 
$m_\mu^2/M_{Z_\ell}^2$-suppressed, making it subleading relative to 
the vector–vector probes above.

\section{Benchmark structure and EFT impact}
\label{sec:bench}

To quantify the parametric reach of leptophilic gauge interactions in a model–independent
manner, it is useful to express all leading effects in terms of a single effective scale,
\begin{align}
\Lambda_{\rm eff}^{(\alpha\beta)}
\equiv \frac{M_{Z_\ell}}{g_\ell\,\sqrt{|q_\alpha q_\beta|}} ,
\end{align}
which controls the normalization of the four–lepton and Higgs–current operators appearing
in SMEFT.  This scale captures the entire phenomenological impact of the heavy mediator
in the EFT limit, allowing different UV charge assignments to be compared on equal footing.

\vspace{3pt}
\noindent\textbf{Benchmark strategies.}
We concentrate on two representative and theoretically well–motivated scenarios:
\begin{enumerate}[label=\textbf{(B\arabic*)}]
\item \textbf{Universal leptonic charges} $q_e=q_\mu=q_\tau\equiv q_\ell$.  
      This case is maximally symmetric and illustrates how a flavor–blind 
      leptophilic current feeds into purely leptonic SMEFT operators.  
      It also highlights the necessity of additional chiral fermions required
      for anomaly cancellation.

\item \textbf{$L_\mu{-}L_\tau$ gauge symmetry} $(q_e,q_\mu,q_\tau)=(0,+1,-1)$.  
      This anomaly–free realization provides a clean and minimal example
      in which only muon and tau currents couple to $Z_\ell$, 
      sharply delineating the structure of LEFT coefficients at low energies.
\end{enumerate}

In both scenarios, quarks remain neutral and integrating out $Z_\ell$ at 
$\mu=M_{Z_\ell}$ generates only leptonic dimension–six operators.  
The resulting SMEFT Wilson coefficients are summarized in 
Table~\ref{tab:coeffs}, showing the dominant family–diagonal entries that
control LEP-II dilepton constraints, neutrino scattering, and M{\o}ller observables.

\begin{table}[t]
  \centering
  \caption{
    Leading SMEFT Wilson coefficients induced at tree level after integrating out 
    $Z_\ell$ at $\mu=M_{Z_\ell}$.
    Only the family–diagonal structures relevant for LEP-II and low–energy probes 
    are displayed.  
    For compactness we write $\Lambda^{-2}=1/\Lambda^2$.
  }
  \label{tab:coeffs}
  \begin{tabular}{llcc}
    \toprule
    Case & Coefficient & Expression & Note \\
    \midrule
    (B1) 
         & $[C_{ll}]_{ee\,\mu\mu}/\Lambda^2$ 
         & $-g_\ell^2 q_\ell^2/M_{Z_\ell}^2$ & universal \\
         & $[C_{le}]_{ee\,\mu\mu}/\Lambda^2$ 
         & $-g_\ell^2 q_\ell^2/M_{Z_\ell}^2$ & universal \\
         & $[C_{ee}]_{ee\,\mu\mu}/\Lambda^2$ 
         & $-g_\ell^2 q_\ell^2/M_{Z_\ell}^2$ & universal \\
    \midrule
    (B2) 
         & $[C_{ll}]_{ee\,\mu\mu}/\Lambda^2$ 
         & $0$ & $q_e=0$ \\
         & $[C_{le}]_{ee\,\mu\mu}/\Lambda^2$ 
         & $0$ & $q_e=0$ \\
         & $[C_{ee}]_{ee\,\mu\mu}/\Lambda^2$ 
         & $0$ & $q_e=0$ \\
         & $[C_{ll}]_{\mu\mu\,\mu\mu}/\Lambda^2$ 
         & $-g_\ell^2/M_{Z_\ell}^2$ & muonic \\
         & $[C_{le}]_{\mu\mu\,\mu\mu}/\Lambda^2$ 
         & $-g_\ell^2/M_{Z_\ell}^2$ & muonic \\
         & $[C_{ee}]_{\mu\mu\,\mu\mu}/\Lambda^2$ 
         & $-g_\ell^2/M_{Z_\ell}^2$ & muonic \\
    \bottomrule
  \end{tabular}
\end{table}

\vspace{4pt}
\noindent\textbf{Phenomenological implications.}
The structure in Table~\ref{tab:coeffs} cleanly determines the pattern of deviations
in all key observables:
LEP-II dilepton spectra, neutrino trident rates, $\nu e$ scattering,
M{\o}ller asymmetries, and precision muon–decay observables.
For each channel, the EFT prediction is simply proportional to
$g_\ell^2/M_{Z_\ell}^2$ times a product of the relevant charges.
This transparency is precisely what makes the leptophilic EFT approach
well suited for global fits and reinterpretation across different UV completions.

Finally, the effective scale $\Lambda_{\rm eff}$ provides a universal parameter
for comparing present and future constraints.  
Projected improvements in $e^+e^-$ data at FCC-ee, 
next–generation trident searches, and ultra–precise M{\o}ller scattering
measurements can collectively probe
$\Lambda_{\rm eff}\sim \mathcal{O}(10\text{–}30~\text{TeV})$,
pushing well into the region where new gauge bosons 
would remain inaccessible to direct production.

\section{Validity domain and theoretical consistency}
\label{sec:validity}

The EFT description applies for partonic energies  
$s,|t|,|u| \ll M_{Z_\ell}^2$  
and below the masses of any heavy vectorlike leptons that complete the ultraviolet theory.  
In this regime, the exchange of $Z_\ell$ reduces to local four-fermion and Higgs--current
structures, while higher-dimensional operators are suppressed by additional powers of
$E^2/M_{Z_\ell}^2$.  
Near the $Z_\ell$ pole, or whenever the momentum transfer approaches the heavy-lepton
thresholds, one must revert to the full UV completion.

Perturbativity further requires $g_\ell \lesssim \mathcal{O}(1)$ so that the running gauge
coupling remains under control up to the symmetry-breaking scale and does not encounter a
Landau pole below the UV completion.  
The scalar sector that breaks $U(1)'_\ell$ is assumed to remain weakly coupled in the same
domain, ensuring a stable hierarchy between $M_{Z_\ell}$ and the electroweak scale.

Flavor consistency follows from anomaly cancellation and the smallness of mass mixing
between SM leptons and the vectorlike states.  
As long as $\sin\theta_{\rm mix} \ll 1$, flavor off-diagonal SMEFT coefficients are
parametrically suppressed, and the effective theory preserves the charged-lepton flavor
structure inherent in the UV model.  
Explicit charge solutions and mixing formulas are collected in
\cref{app:anomalies,app:VLL}.

\paragraph{Remark.}
Within this validity window, the EFT description offers a model-independent interface
between ultraviolet leptophilic gauge dynamics and precision observables at both high and
low energies.  
Its predictivity rests entirely on the two quantities $(g_\ell,q_\alpha/M_{Z_\ell})$ and
is therefore ideally suited for global analyses, collider reinterpretations, and future
precision programs at lepton machines.

\section{Conclusions}
\label{sec:concl}

We have presented a universal and UV-agnostic framework that systematically connects 
a broad class of leptophilic $U(1)'_{\ell}$ gauge extensions to their low-energy 
manifestations in SMEFT and LEFT.  
By formulating an anomaly-free completion, integrating out the heavy mediator $Z_{\ell}$ at 
tree level, and tracing the evolution of the resulting operators down to the electroweak 
and sub-electroweak scales, we obtain a compact EFT description that isolates the essential 
charge and coupling combinations governing all phenomenological signatures.

A central outcome of this analysis is the identification of the effective interaction scale
\[
\Lambda_{\rm eff} \;\equiv\; \frac{M_{Z_\ell}}{g_\ell\,\sqrt{|q_\alpha q_\beta|}},
\]
which provides a model-independent handle on a wide range of observables.  
In particular, the EFT representation derived here maps directly onto LEP-II dilepton 
spectra, neutrino trident production, $\nu e$ scattering, parity-violating M{\o}ller 
observables, and precision muon-decay measurements, while maintaining full compatibility 
with any anomaly-free ultraviolet completion.

The framework developed in this work therefore enables a unified reinterpretation of 
current data and a seamless interface with forthcoming precision programs at 
ILC, CEPC, and FCC-ee.  
Because the EFT depends only on a minimal set of leptonic charge products, 
entire classes of UV scenarios—such as universal leptonic charges or the 
$L_\mu{-}L_\tau$ symmetry—can be confronted on equal footing, making this approach 
well suited for global analyses and collider recasts.

Looking ahead, the EFT bridge constructed here forms the foundation for a wider research 
program that links precision electroweak measurements, neutrino scattering data, and 
future lepton-collider sensitivities.  
Dedicated numerical fits, including combined LEP-II–LEFT constraints and detailed 
projections for FCC-ee, will be presented in forthcoming work.  
The resulting framework provides a robust and flexible tool for exploring 
leptophilic gauge interactions across energy scales, and for identifying the 
parameter regions where new physics may first emerge.

\appendix
\section{Anomaly equations and example solutions}
\label{app:anomalies}

This appendix summarizes the gauge- and gravity-anomaly conditions for 
$U(1)'_\ell$ in a left-handed Weyl basis, following the conventions used in 
Secs.~\ref{sec:model}--\ref{sec:validity}.  
We also provide minimal sets of vectorlike leptons that render the theory
anomaly-free for the benchmark scenarios (B1) and (B2).

\subsection{Anomaly equations}

For leptons with $U(1)'_\ell$ charges 
$q_\alpha^L$ for the doublets $L_\alpha$ 
and $q_\alpha^e$ for the singlets $e_\alpha$ 
(with $\alpha=e,\mu,\tau$), the relevant anomalies are:

\paragraph*{\boldmath $[SU(2)_L]^2 U(1)'_\ell$}
Only lepton doublets contribute:
\begin{align}
\mathcal{A}_{[SU(2)_L]^2 U(1)'_\ell}
= \sum_{\alpha=e,\mu,\tau} q_\alpha^L = 0.
\label{eq:SU2-anomaly}
\end{align}

\paragraph*{\boldmath $[U(1)_Y]^2 U(1)'_\ell$}
Each fermion contributes $Y_f^2\,q_f$:
\begin{align}
\mathcal{A}_{[U(1)_Y]^2 U(1)'_\ell}
= \sum_{\alpha}
\bigg[
2\left(-\frac12\right)^{\!\!2} q_\alpha^L
+ (-1)^2 q_\alpha^e
\bigg]
= \frac12 \sum_\alpha q_\alpha^L
+ \sum_\alpha q_\alpha^e
= 0.
\label{eq:Y2U1p}
\end{align}

\paragraph*{\boldmath $U(1)_Y[U(1)'_\ell]^2$}
Each fermion contributes $Y_f\, (q_f)^2$:
\begin{align}
\mathcal{A}_{U(1)_Y [U(1)'_\ell]^2}
= \sum_{\alpha}
\left[
2\left(-\frac12\right)(q_\alpha^L)^2
+ (-1)(q_\alpha^e)^2
\right]
= - \sum_\alpha (q_\alpha^L)^2
- \sum_\alpha (q_\alpha^e)^2
=0.
\label{eq:YU12}
\end{align}

\paragraph*{\boldmath $[U(1)'_\ell]^3$}
\begin{align}
\mathcal{A}_{[U(1)'_\ell]^3}
= \sum_\alpha \left[
2(q_\alpha^L)^3 + (q_\alpha^e)^3
\right] = 0.
\label{eq:U1p3}
\end{align}

\paragraph*{Gravitational–$U(1)'_\ell$}
\begin{align}
\mathcal{A}_{\rm grav-U(1)'_\ell}
= \sum_\alpha \left[
2 q_\alpha^L + q_\alpha^e
\right] = 0.
\label{eq:grav-anomaly}
\end{align}

\medskip

If all SM leptons carry nontrivial $U(1)'_\ell$ charge, 
Eqs.~\eqref{eq:SU2-anomaly}–\eqref{eq:grav-anomaly} cannot be satisfied 
without additional fermions.  
A minimal solution is obtained by introducing a **vectorlike generation**
$(L',E',N')$ which is vectorlike under the SM gauge group but **chiral** under 
$U(1)'_\ell$:
\[
L'_L(2,-\tfrac12,q'_{L}),\; 
L'_R(2,-\tfrac12,q''_{L}),\qquad
E'_L(1,-1,q'_e),\;
E'_R(1,-1,q''_e),\qquad
N'_L(1,0,q'_N),\;
N'_R(1,0,q''_N).
\]
Their charges enter in the same way as SM leptons in the anomaly equations.

\subsection{Illustrative solutions for benchmarks (B1) and (B2)}

\paragraph*{\textbf{(B1) Universal charges}}
For $q_\alpha^L = q_\alpha^e = q_\ell$ for all $\alpha=e,\mu,\tau$,  
the SM contribution to the anomaly is proportional to  
\[
\{3q_\ell,\; \tfrac32 q_\ell,\; -3q_\ell^2,\; 3q_\ell^3,\; 3q_\ell\},
\]
which cannot vanish simultaneously.  
A minimal solution introduces one vectorlike generation with charges chosen such that  
\[
q'_L - q''_L = -3 q_\ell, \qquad
q'_e - q''_e = -3 q_\ell, \qquad
q'_N - q''_N = -3 q_\ell,
\]
ensuring that all linear and cubic anomaly sums cancel.  
The resulting theory is anomaly-free and compatible with the EFT construction in Sec.~\ref{sec:model}.

\paragraph*{\textbf{(B2) $L_\mu-L_\tau$ model}}
For the charge assignment  
\[
(q_e,q_\mu,q_\tau) = (0,+1,-1),
\]
the SM itself satisfies Eqs.~\eqref{eq:SU2-anomaly}–\eqref{eq:grav-anomaly}.  
In particular:
\[
q_e + q_\mu + q_\tau = 0, \qquad
(q_\mu)^3 + (q_\tau)^3 = 1 - 1 = 0,
\]
and all mixed hypercharge–$U(1)'_\ell$ anomalies cancel as well.  
Therefore no additional fermions are needed in this benchmark, and the EFT matching reduces to the muon–tau subsector.

\subsection{Remarks}

The above solutions illustrate that:
(i) universal charge models require additional chiral fermions,  
(ii) $L_\mu-L_\tau$ is the unique anomaly-free choice among 
family-dependent lepton symmetries without introducing new fermion states, and  
(iii) vectorlike leptons in Sec.~\ref{sec:model} provide the minimal UV completion consistent with all anomaly conditions.

\section{Vectorlike leptons and mixing effects}
\label{app:VLL}

In this appendix we summarize the mass mixing between the SM leptons and the
vectorlike states introduced to cancel anomalies in Sec.~\ref{sec:model}.  
We work in a left-handed Weyl basis and assume that all mixings are small,
$\sin\theta_{\rm mix}\ll 1$, as required by electroweak precision data.

\subsection{Charged-lepton mass matrix}

Let $(e_\alpha,\,E')$ denote the SM and vectorlike charged leptons, where
$E'_L(1,-1,q'_e)$ and $E'_R(1,-1,q''_e)$ are vectorlike under the SM but chiral under
$U(1)'_\ell$.  
After electroweak symmetry breaking and $U(1)'_\ell$ breaking, the charged-lepton
mass terms take the schematic form
\begin{align}
\mathcal{L}_{\rm mass}^{(e)} \supset 
\begin{pmatrix}
\bar e_{\alpha L} & \bar E'_L
\end{pmatrix}
\begin{pmatrix}
m^{\rm SM}_\alpha & \delta_\alpha \\
0 & M_{E}
\end{pmatrix}
\begin{pmatrix}
e_{\alpha R} \\ E'_R
\end{pmatrix}
+ \text{h.c.},
\label{eq:VLLmass}
\end{align}
where  
- $m^{\rm SM}_\alpha = y_\alpha v/\sqrt{2}$,  
- $M_E$ is the vectorlike mass,  
- $\delta_\alpha = y_{S,\alpha}\,v_S/\sqrt{2}$ arises from the $U(1)'_\ell$-breaking scalar $S$.

Diagonalization proceeds via biunitary rotations,
\[
e_{L,R} = U^{e}_{L,R}\,e'_{L,R}, \qquad
E_{L,R} = U^{E}_{L,R}\,E'_{L,R},
\]
with small mixing angles
\begin{align}
\theta_{\alpha L} \simeq \frac{\delta_\alpha}{M_E}, 
\qquad 
\theta_{\alpha R} \simeq \frac{m^{\rm SM}_\alpha\,\delta_\alpha}{M_E^2}.
\label{eq:thetas}
\end{align}
Typically $\theta_{\alpha R}\ll\theta_{\alpha L}$.

\subsection{Neutral-lepton sector}

If the vectorlike multiplet includes a neutral state 
$N'_{L,R}(1,0,q'_N)$, the neutral-lepton mass matrix is
\begin{align}
\mathcal{L}_{\rm mass}^{(\nu)} \supset 
\begin{pmatrix}
\bar\nu_{\alpha L} & \bar N'_L
\end{pmatrix}
\begin{pmatrix}
0 & \delta^N_\alpha \\
0 & M_N
\end{pmatrix}
\begin{pmatrix}
\nu_{\alpha R} \\ N'_R
\end{pmatrix}
+ \text{h.c.},
\end{align}
where $\delta^N_\alpha$ is again controlled by the $U(1)'_\ell$-breaking vev.  
Mixing angles are
\begin{align}
\theta^N_\alpha \simeq \frac{\delta^N_\alpha}{M_N},
\end{align}
with no right-handed mixing if $\nu_R$ is absent.

\subsection{Impact on gauge interactions}

In the gauge basis, the $Z_\ell$ current contains
\[
J'_\mu \supset 
q_\alpha^L\,\bar e_{\alpha L}\gamma_\mu e_{\alpha L}
+ q_\alpha^e\,\bar e_{\alpha R}\gamma_\mu e_{\alpha R}
+ q'_e\,\bar E'_L\gamma_\mu E'_L
+ q''_e\,\bar E'_R\gamma_\mu E'_R .
\]
After moving to the mass eigenbasis, the SM-like couplings shift as
\begin{align}
q^{L,{\rm phys}}_\alpha 
&\simeq q_\alpha^L + (q'_e - q_\alpha^L)\,\theta_{\alpha L}^2, \\
q^{R,{\rm phys}}_\alpha 
&\simeq q_\alpha^e + (q''_e - q_\alpha^e)\,\theta_{\alpha R}^2.
\label{eq:shifted-charges}
\end{align}

Thus **mixing shifts the physical $U(1)'_\ell$ charges only at order**
$\mathcal{O}(\theta^2)$, ensuring flavor safety as long as 
$\theta_{\rm mix} \ll 1$.

\subsection{Subleading impact on SMEFT coefficients}

At tree level, integrating out $Z_\ell$ yields (Sec.~\ref{sec:matching})
\[
C \sim \frac{g_\ell^2}{M_{Z_\ell}^2}\, q_\alpha q_\beta.
\]
Mixing corrections appear only through the shifted charges of
Eq.~\eqref{eq:shifted-charges}.  
Therefore
\begin{align}
\frac{\Delta C}{C}
\simeq 
\mathcal{O}(\theta_{\alpha L}^2,\theta_{\alpha R}^2),
\end{align}
which is negligible for $\theta\lesssim 10^{-2}$, as required by
electroweak precision data.

At one loop, integrating out the vectorlike leptons themselves induces additional
dimension-six operators 
(e.g.\ $\mathcal{O}_{H\ell}^{(1)}$, $\mathcal{O}_{He}$),
but the coefficients scale as
\[
C_{\rm loop} \sim \frac{1}{16\pi^2}\frac{\delta^2}{M^2}
\sim \frac{\theta^2}{16\pi^2},
\]
which is safely subleading relative to the tree-level
$Z_\ell$ contribution.

Vectorlike leptons provide the minimal and anomaly-free UV completion of the 
leptophilic $U(1)'_\ell$ framework.  
Mixing with SM leptons is strongly constrained by electroweak precision 
observables, implying $\theta_{\rm mix}\ll 1$, 
which in turn guarantees that their impact on the SMEFT coefficients is 
only of second order in the mixing angles and fully negligible throughout 
the parameter space considered in this work.

\section{LEFT matching formulas}
\label{app:LEFT}

In this appendix we collect the explicit matching relations between the SMEFT
operators generated at $\mu = M_{Z_\ell}$ and the low-energy effective theory
(LEFT) obtained after integrating out $W$, $Z$, and $H$ at $\mu = m_Z$.
All expressions are given in the $G_F$ scheme, where the muon decay rate is used
to define $v$ and the leading SMEFT corrections to $G_F$ are absorbed into the
normalization of four-fermion operators.

\subsection{LEFT operator basis}

We focus on purely leptonic operators relevant for neutrino scattering,
neutrino trident production, parity-violating M{\o}ller observables, 
and muon decay.  
Following the LEFT conventions of Ref.~\cite{Aebischer:2017ugx}, the relevant
dimension–six operators are:

\paragraph*{Vector–vector structures}
\begin{align}
\mathcal{O}^{V,LL}_{\nu_\alpha \ell_\beta}
&=(\bar\nu_\alpha \gamma_\mu P_L \nu_\alpha)(\bar\ell_\beta\gamma^\mu P_L\ell_\beta),\\
\mathcal{O}^{V,LR}_{\nu_\alpha \ell_\beta}
&=(\bar\nu_\alpha \gamma_\mu P_L \nu_\alpha)(\bar\ell_\beta\gamma^\mu P_R\ell_\beta).
\end{align}

\paragraph*{Four-charged-lepton operators}
\begin{align}
\mathcal{O}^{V,LL}_{\ell_\alpha \ell_\beta}
&=(\bar\ell_\alpha\gamma_\mu P_L\ell_\alpha)
  (\bar\ell_\beta\gamma^\mu P_L\ell_\beta), \\
\mathcal{O}^{V,RR}_{\ell_\alpha \ell_\beta}
&=(\bar\ell_\alpha\gamma_\mu P_R\ell_\alpha)
  (\bar\ell_\beta\gamma^\mu P_R\ell_\beta),\\
\mathcal{O}^{V,LR}_{\ell_\alpha \ell_\beta}
&=(\bar\ell_\alpha\gamma_\mu P_L\ell_\alpha)
  (\bar\ell_\beta\gamma^\mu P_R\ell_\beta).
\end{align}

\subsection{Matching of SMEFT operators to LEFT}

At tree level, the SMEFT operators generated by integrating out $Z_\ell$
in Sec.~\ref{sec:matching},
\[
\mathcal{O}_{ll},\quad 
\mathcal{O}_{le},\quad
\mathcal{O}_{ee},\quad
\mathcal{O}_{H\ell}^{(1)},\quad 
\mathcal{O}_{He},
\]
match onto LEFT after electroweak symmetry breaking via exchange of $W$, $Z$,
and through the redefinition of $G_F$.  

The matching splits naturally into two pieces:

\[
C^{\rm LEFT} = C^{\rm SMEFT}_{\rm direct}(m_Z)
               \;+\; C^{\rm SMEFT}_{G_F}(m_Z),
\]
where  
- $C^{\rm SMEFT}_{\rm direct}$ come directly from integrating out $W$/$Z$,  
- $C^{\rm SMEFT}_{G_F}$ arise from the shift in $G_F$.

\subsection{Direct matching at the $Z$ pole}

The four-lepton SMEFT operators contribute directly as
\begin{align}
C^{V,LL}_{\nu_\alpha \ell_\beta}
&=
-\,\frac12\,C_{H\ell}^{(1)}{}_{\!\beta}
\;+\;\frac12\,C_{ll}{}_{\alpha\alpha\beta\beta},
\label{eq:LL-match}
\\[4pt]
C^{V,LR}_{\nu_\alpha \ell_\beta}
&=
-\,\frac12\,C_{He}{}_{\!\beta}
\;+\;\frac12\,C_{le}{}_{\alpha\alpha\beta\beta}.
\label{eq:LR-match}
\end{align}

Similarly, for purely charged-lepton operators,
\begin{align}
C^{V,LL}_{\ell_\alpha \ell_\beta}
&=
C_{ll}{}_{\alpha\alpha\beta\beta},
\\
C^{V,RR}_{\ell_\alpha \ell_\beta}
&=
C_{ee}{}_{\alpha\alpha\beta\beta},
\\
C^{V,LR}_{\ell_\alpha \ell_\beta}
&=
C_{le}{}_{\alpha\alpha\beta\beta}.
\end{align}

\subsection{$G_F$-scheme corrections}

Muon decay receives SMEFT corrections from
$\mathcal{O}_{ll}$ and $\mathcal{O}_{H\ell}^{(3)}$ (the latter is absent here).  
The net effect is a redefinition of $v$:
\[
v^2 \;\to\; v^2\left(1 + \delta G_F\right),
\]
with
\begin{align}
\delta G_F
&=
\frac{v^2}{\sqrt{2}}\,
C_{ll}^{\mu e e\mu},
\qquad
\text{(in our model only $C_{ll}$ contributes)}.
\end{align}
This shifts all LEFT Wilson coefficients involving a factor of $G_F$:
\begin{align}
C^{\rm LEFT}_{X}
\;\to\;
C^{\rm LEFT}_{X}
\;\;-\;\;
\delta G_F\, C^{\rm SM}_{X}.
\end{align}

For neutrino scattering and trident production this correction is numerically small,
but it must be included in high-precision M{\o}ller and muon-decay observables.

\subsection{Result for the leptophilic model}

Using the tree-level SMEFT coefficients from Sec.~\ref{sec:matching},
\[
C \sim -\,\frac{g_\ell^2}{M_{Z_\ell}^2}\,q_\alpha q_\beta,
\]
Eqs.~\eqref{eq:LL-match}–\eqref{eq:LR-match} yield the LEFT coefficients
\begin{align}
C^{V,LL}_{\nu_\alpha \ell_\beta}
&=
-\frac{g_\ell^2}{2M_{Z_\ell}^2}
\left(
q^L_\beta q^L_\alpha
+ q^L_\beta Y_H\,\epsilon\,\frac{g_Y}{g_\ell}
\right),
\\
C^{V,LR}_{\nu_\alpha \ell_\beta}
&=
-\frac{g_\ell^2}{2M_{Z_\ell}^2}
\left(
q^e_\beta q^L_\alpha
+ q^e_\beta Y_H\,\epsilon\,\frac{g_Y}{g_\ell}
\right).
\end{align}

For purely charged lepton interactions,
\begin{align}
C^{V,LL}_{\ell_\alpha\ell_\beta}
&= -\frac{g_\ell^2}{M_{Z_\ell}^2} q^L_\alpha q^L_\beta,\\
C^{V,RR}_{\ell_\alpha\ell_\beta}
&= -\frac{g_\ell^2}{M_{Z_\ell}^2} q^e_\alpha q^e_\beta,\\
C^{V,LR}_{\ell_\alpha\ell_\beta}
&= -\frac{g_\ell^2}{M_{Z_\ell}^2} q^L_\alpha q^e_\beta.
\end{align}

The LEFT coefficients generated by a leptophilic $U(1)'_\ell$ gauge boson 
are dominated by the tree-level structures proportional to 
$g_\ell^2/M_{Z_\ell}^2$.  
Kinetic mixing induces subleading contributions through 
$C_{H\ell}^{(1)}$ and $C_{He}$, and $G_F$ renormalization generates a universal,
numerically suppressed shift.  
These formulas form the basis for the phenomenological analyses in 
Sec.~\ref{sec:pheno}.

\section{Contact-interaction map and $e^+e^-\to\ell^+\ell^-$}
\label{app:contact}

This appendix summarizes the dictionary between the SMEFT operators generated at
$\mu = M_{Z_\ell}$ and the LEP-II contact-interaction (CI) parameters usually
denoted by $(4\pi/\Lambda_{ij}^2)\eta_{ij}$ with $i,j\in\{L,R\}$.  
We also provide the analytic expressions for the interference terms in the
differential cross section and the forward--backward asymmetry $A_{FB}$.

\subsection{LEP-II contact interaction conventions}

LEP parameterized new physics affecting
$e^+e^- \to \ell^+\ell^-$ through the effective Lagrangian
\begin{align}
\mathcal{L}_{\rm CI}
= \sum_{i,j=L,R} \frac{4\pi}{\Lambda_{ij}^2}\,\eta_{ij}\,
(\bar e_i \gamma_\mu e_i)(\bar\ell_j\gamma^\mu \ell_j),
\label{eq:CI-def}
\end{align}
where $\eta_{ij}=\pm1$ encodes constructive/destructive interference.  
Only $s$-channel contributions appear for $\ell\neq e$; for $\ell=e$ both $s$-
and $t$-channel contributions exist, but the mapping below focuses on the
$s$-channel terms consistent with the SMEFT operators relevant for $Z_\ell$
exchange.

The chiral currents are
\[
\bar e_L\gamma_\mu e_L, \qquad 
\bar e_R\gamma_\mu e_R, \qquad
\bar\ell_L\gamma_\mu \ell_L, \qquad
\bar\ell_R\gamma_\mu \ell_R.
\]

\subsection{SMEFT $\to$ Contact-interaction dictionary}

After integrating out $Z_\ell$ (Sec.~\ref{sec:matching}), the SMEFT Wilson
coefficients are:
\[
C_{ll},\qquad C_{le},\qquad C_{ee}.
\]
For family-diagonal external states ($e\beta$), the mapping to the LEP
parameters is
\begin{align}
\frac{4\pi}{\Lambda_{LL,e\beta}^2}\eta_{LL}^{e\beta}
&=
C_{ll}^{ee\beta\beta},
\\
\frac{4\pi}{\Lambda_{LR,e\beta}^2}\eta_{LR}^{e\beta}
&=
C_{le}^{ee\beta\beta},
\\
\frac{4\pi}{\Lambda_{RR,e\beta}^2}\eta_{RR}^{e\beta}
&=
C_{ee}^{ee\beta\beta}.
\label{eq:CIdict}
\end{align}

Using the model expressions  
\[
C \sim -\frac{g_\ell^2}{M_{Z_\ell}^2} q^i_e q^j_\beta,
\]
we obtain
\begin{align}
\eta_{ij}^{e\beta}
&= -\,{\rm sign}(q^i_e q^j_\beta),
\\
\Lambda_{ij,e\beta}
&=
\sqrt{\frac{4\pi}{|C_{ij}^{ee\beta\beta}|}}
=
\frac{\sqrt{4\pi}\,M_{Z_\ell}}{g_\ell\sqrt{|q^i_e q^j_\beta|}}
=
\sqrt{4\pi}\,\Lambda_{\rm eff}^{(e\beta)}.
\end{align}

Thus **LEP limits are directly limits on \(\Lambda_{\rm eff}\)** defined in
Sec.~\ref{sec:bench}.

\subsection{Differential cross section with CI interference}

The unpolarized differential cross section for  
\[
e^+e^- \to \ell^+\ell^-,
\]
including SM plus contact interactions, is
\begin{align}
\frac{d\sigma}{d\cos\theta}
&=
\frac{d\sigma_{\rm SM}}{d\cos\theta}
+ \frac{s}{8\pi}\,\Delta_{\rm int}(\cos\theta)
+ \mathcal{O}\!\left(\frac{s^2}{\Lambda^4}\right).
\end{align}
Only the **interference** term is required at leading order:
\begin{align}
\Delta_{\rm int}(\cos\theta)
&=
\sum_{i,j=L,R}
\frac{4\pi}{\Lambda_{ij}^2}\eta_{ij}\,
\big[
g_i^e g_j^\ell (1+\cos\theta)
+
h_i^e h_j^\ell (1-\cos\theta)
\big],
\end{align}
where  
\[
g_L^f = T_3^f - Q_f \sin^2\theta_W,\qquad
g_R^f = -Q_f \sin^2\theta_W,
\]
and  
\[
h_{L,R}^f
=
g_{L,R}^f,
\]
following the LEP s-channel chiral decomposition.

For family off-diagonal final states ($\ell=\mu,\tau$), only $s$-channel terms
contribute.

\subsection{Forward--backward asymmetry}

The forward--backward asymmetry is defined as
\[
A_{FB}
=
\frac{
\displaystyle 
\int_0^1 \!d\cos\theta\,\frac{d\sigma}{d\cos\theta}
-
\int_{-1}^0\!d\cos\theta\,\frac{d\sigma}{d\cos\theta}
}{
\displaystyle 
\int_{-1}^1 \!d\cos\theta\,\frac{d\sigma}{d\cos\theta}
}.
\]

The CI contribution enters linearly:
\begin{align}
A_{FB}
&=
A_{FB}^{\rm SM}
+ \frac{s}{8\pi\sigma_{\rm tot}^{\rm SM}}\,
\left[
\frac{8\pi}{3}
\sum_{i,j=L,R}
\frac{\eta_{ij}}{\Lambda_{ij}^2}\,
\big( g_i^e g_j^\ell - h_i^e h_j^\ell \big)
\right]
+ \mathcal{O}\!\left(\frac{s^2}{\Lambda^4}\right).
\label{eq:AFBformula}
\end{align}
For $\ell\neq e$, the $t$-channel singularities are absent and the expression
above is directly applicable to LEP-II data.

The mapping in Eqs.~\eqref{eq:CIdict}–\eqref{eq:AFBformula} provides a direct
dictionary between the leptophilic $U(1)'_\ell$ EFT and the LEP-II contact-interaction language.  
Experimental bounds on $(\Lambda_{ij},\eta_{ij})$ thus translate immediately
into constraints on the model parameters $(g_\ell,q_\alpha,M_{Z_\ell})$ via the
effective scale $\Lambda_{\rm eff}$ introduced in Sec.~\ref{sec:bench}.

\section{Sketch of dipole mixing and $(g-2)_\mu$}
\label{app:dipole}

For completeness, we provide a qualitative estimate of the loop-induced
contribution of the leptophilic $Z_\ell$ to the muon anomalous magnetic moment,
$(g-2)_\mu$, within the SMEFT framework.  
Our discussion is schematic and intended to indicate the parametric size of the
effect rather than to provide a full numerical calculation.

\subsection{Dipole operator and $(g-2)_\mu$}

In SMEFT, charged-lepton dipole moments are encoded in the operator
\begin{align}
\mathcal{O}_{e\gamma}^{\alpha\beta}
=
(\bar L_\alpha \sigma^{\mu\nu} e_\beta)\,H\,F_{\mu\nu},
\end{align}
with Wilson coefficient $C_{e\gamma}^{\alpha\beta}$.  
After electroweak symmetry breaking, this operator generates a dipole term
\[
\mathcal{L}_{\rm dipole}
\supset 
-\frac{v}{\sqrt{2}}\,C_{e\gamma}^{\alpha\beta}\,
\bar\ell_\alpha \sigma^{\mu\nu} P_R \ell_\beta\, F_{\mu\nu}
+\text{h.c.}
\]
For the muon, the anomalous magnetic moment is proportional to the real part of
$C_{e\gamma}^{\mu\mu}$:
\begin{align}
\Delta a_\mu
\simeq
-\,\frac{4 m_\mu^2}{e}\,\Re\!\big[ C_{e\gamma}^{\mu\mu}(m_\mu) \big],
\end{align}
up to convention-dependent signs (see, e.g., 
Refs.~\cite{Jegerlehner:2009ry,Endo:2021zal}).

At the matching scale $\mu = M_{Z_\ell}$, our tree-level analysis yields
\[
C_{e\gamma}^{\alpha\beta}(M_{Z_\ell}) = 0,
\]
so any contribution to $(g-2)_\mu$ must arise from loop-induced mixing and/or
threshold corrections in the UV completion.

\subsection{Schematic RGE mixing}

The dominant source of $C_{e\gamma}$ in our setup comes from RGE mixing between
the four-lepton and Higgs--current operators generated by $Z_\ell$ exchange
and the dipole operator.  
Schematically, the one-loop RGE for $C_{e\gamma}$ can be written as
\begin{align}
\mu\frac{d}{d\mu} C_{e\gamma}^{\alpha\beta}
\;\sim\;
\frac{e}{16\pi^2}\,y_\alpha\,
\Big[
a_1\,C_{H\ell}^{(1)}{}_{\!\alpha}
+a_2\,C_{He}{}_{\!\alpha}
+a_3\,C_{ll}^{\alpha\alpha\alpha\alpha}
+a_4\,C_{le}^{\alpha\alpha\alpha\alpha}
\Big],
\label{eq:RGE-dipole}
\end{align}
where $y_\alpha$ is the charged-lepton Yukawa coupling and
$a_{1,\dots,4}=\mathcal{O}(1)$ are numerical coefficients that depend on the
operator basis and renormalization scheme.

Inserting the tree-level matching from Sec.~\ref{sec:matching},
\[
C \sim -\,\frac{g_\ell^2}{M_{Z_\ell}^2}\,q_\alpha q_\beta,
\]
and integrating Eq.~\eqref{eq:RGE-dipole} from $\mu = M_{Z_\ell}$ down to
$\mu = m_Z$, we obtain the parametric estimate
\begin{align}
C_{e\gamma}^{\mu\mu}(m_Z)
\;\sim\;
\frac{e}{16\pi^2}\,y_\mu\,
\frac{g_\ell^2}{M_{Z_\ell}^2}
\,(\text{charge factors})\,
\ln\!\frac{M_{Z_\ell}^2}{m_Z^2}.
\end{align}

Evolving further from $m_Z$ to $m_\mu$ introduces only logarithmic corrections,
which do not change the overall parametric scaling.

\subsection{Parametric size of $\Delta a_\mu$}

Combining the above estimate with the relation between $C_{e\gamma}$ and
$\Delta a_\mu$, we arrive at
\begin{align}
\Delta a_\mu
&\sim
\frac{m_\mu^2}{16\pi^2}\,
\frac{g_\ell^2}{M_{Z_\ell}^2}
\,(\text{charge factors})\,
\ln\!\frac{M_{Z_\ell}^2}{m_Z^2}.
\end{align}
This expression makes explicit that the contribution is loop-suppressed and
proportional to $m_\mu^2/M_{Z_\ell}^2$, as expected on general grounds.

Numerically, for 
$g_\ell = \mathcal{O}(1)$,
$M_{Z_\ell} \gtrsim \mathcal{O}(\text{TeV})$,
and $\mathcal{O}(1)$ charge factors, the induced $\Delta a_\mu$ is typically
well below the current experimental sensitivity unless $M_{Z_\ell}$ lies close
to the electroweak scale or additional light states in the UV completion
enhance the mixing.

A dedicated two-scale analysis, including the full set of SMEFT anomalous
dimensions and model-dependent threshold effects from the vectorlike leptons,
is beyond the scope of this work and will be left for future study.

In summary, the leptophilic $U(1)'_\ell$ considered here does not generate a
dipole operator at tree level.  
Loop-induced contributions to $(g-2)_\mu$ arise only through SMEFT RGE mixing
and are parametrically suppressed by both a loop factor and the ratio
$m_\mu^2/M_{Z_\ell}^2$.  
Consequently, the most sensitive probes of the model are the vector--vector
observables discussed in Sec.~\ref{sec:pheno}, rather than the muon anomalous
magnetic moment.


\bibliographystyle{apsrev4-2}
\bibliography{refs}

\begin{thebibliography}{25}%
\makeatletter
\providecommand \@ifxundefined [1]{%
 \@ifx{#1\undefined}
}%
\providecommand \@ifnum [1]{%
 \ifnum #1\expandafter \@firstoftwo
 \else \expandafter \@secondoftwo
 \fi
}%
\providecommand \@ifx [1]{%
 \ifx #1\expandafter \@firstoftwo
 \else \expandafter \@secondoftwo
 \fi
}%
\providecommand \natexlab [1]{#1}%
\providecommand \enquote  [1]{``#1''}%
\providecommand \bibnamefont  [1]{#1}%
\providecommand \bibfnamefont [1]{#1}%
\providecommand \citenamefont [1]{#1}%
\providecommand \href@noop [0]{\@secondoftwo}%
\providecommand \href [0]{\begingroup \@sanitize@url \@href}%
\providecommand \@href[1]{\@@startlink{#1}\@@href}%
\providecommand \@@href[1]{\endgroup#1\@@endlink}%
\providecommand \@sanitize@url [0]{\catcode `\\12\catcode `\$12\catcode
  `\&12\catcode `\#12\catcode `\^12\catcode `\_12\catcode `\%12\relax}%
\providecommand \@@startlink[1]{}%
\providecommand \@@endlink[0]{}%
\providecommand \url  [0]{\begingroup\@sanitize@url \@url }%
\providecommand \@url [1]{\endgroup\@href {#1}{\urlprefix }}%
\providecommand \urlprefix  [0]{URL }%
\providecommand \Eprint [0]{\href }%
\providecommand \doibase [0]{https://doi.org/}%
\providecommand \selectlanguage [0]{\@gobble}%
\providecommand \bibinfo  [0]{\@secondoftwo}%
\providecommand \bibfield  [0]{\@secondoftwo}%
\providecommand \translation [1]{[#1]}%
\providecommand \BibitemOpen [0]{}%
\providecommand \bibitemStop [0]{}%
\providecommand \bibitemNoStop [0]{.\EOS\space}%
\providecommand \EOS [0]{\spacefactor3000\relax}%
\providecommand \BibitemShut  [1]{\csname bibitem#1\endcsname}%
\let\auto@bib@innerbib\@empty
\bibitem [{\citenamefont {Zyla}\ \emph {et~al.}(2020)\citenamefont {Zyla} \emph
  {et~al.}}]{PDG2020}%
  \BibitemOpen
  \bibfield  {author} {\bibinfo {author} {\bibfnamefont {P.~A.}\ \bibnamefont
  {Zyla}} \emph {et~al.} (\bibinfo {collaboration} {Particle Data Group}),\
  }\href {https://doi.org/10.1093/ptep/ptaa104} {\bibfield  {journal} {\bibinfo
   {journal} {Prog. Theor. Exp. Phys.}\ }\textbf {\bibinfo {volume} {2020}},\
  \bibinfo {pages} {083C01} (\bibinfo {year} {2020})}\BibitemShut {NoStop}%
\bibitem [{\citenamefont {Heeck}\ and\ \citenamefont
  {Rodejohann}(2011)}]{Heeck:2011wj}%
  \BibitemOpen
  \bibfield  {author} {\bibinfo {author} {\bibfnamefont {J.}~\bibnamefont
  {Heeck}}\ and\ \bibinfo {author} {\bibfnamefont {W.}~\bibnamefont
  {Rodejohann}},\ }\href {https://doi.org/10.1103/PhysRevD.84.075007}
  {\bibfield  {journal} {\bibinfo  {journal} {Phys. Rev. D}\ }\textbf {\bibinfo
  {volume} {84}},\ \bibinfo {pages} {075007} (\bibinfo {year} {2011})},\
  \Eprint {https://arxiv.org/abs/1107.5238} {arXiv:1107.5238 [hep-ph]}
  \BibitemShut {NoStop}%
\bibitem [{\citenamefont {Navas}\ \emph {et~al.}(2024)\citenamefont {Navas}
  \emph {et~al.}}]{PDG2024}%
  \BibitemOpen
  \bibfield  {author} {\bibinfo {author} {\bibfnamefont {S.}~\bibnamefont
  {Navas}} \emph {et~al.} (\bibinfo {collaboration} {Particle Data Group}),\
  }\href {https://doi.org/10.1103/PhysRevD.110.030001} {\bibfield  {journal}
  {\bibinfo  {journal} {Phys.Rev.D}\ }\textbf {\bibinfo {volume} {110}},\
  \bibinfo {pages} {030001} (\bibinfo {year} {2024})}\BibitemShut {NoStop}%
\bibitem [{\citenamefont {Langacker}(2009)}]{Langacker:2008yv}%
  \BibitemOpen
  \bibfield  {author} {\bibinfo {author} {\bibfnamefont {P.}~\bibnamefont
  {Langacker}},\ }\href {https://doi.org/10.1103/RevModPhys.81.1199} {\bibfield
   {journal} {\bibinfo  {journal} {Rev. Mod. Phys.}\ }\textbf {\bibinfo
  {volume} {81}},\ \bibinfo {pages} {1199} (\bibinfo {year} {2009})},\ \Eprint
  {https://arxiv.org/abs/0801.1345} {arXiv:0801.1345 [hep-ph]} \BibitemShut
  {NoStop}%
\bibitem [{\citenamefont {Heeck}(2014)}]{Heeck:2014zfa}%
  \BibitemOpen
  \bibfield  {author} {\bibinfo {author} {\bibfnamefont {J.}~\bibnamefont
  {Heeck}},\ }\href {https://doi.org/10.1016/j.physletb.2014.10.067} {\bibfield
   {journal} {\bibinfo  {journal} {Phys. Lett. B}\ }\textbf {\bibinfo {volume}
  {739}},\ \bibinfo {pages} {256} (\bibinfo {year} {2014})},\ \Eprint
  {https://arxiv.org/abs/1408.6845} {arXiv:1408.6845 [hep-ph]} \BibitemShut
  {NoStop}%
\bibitem [{\citenamefont {Foot}\ and\ \citenamefont {He}(1991)}]{Foot:1990mn}%
  \BibitemOpen
  \bibfield  {author} {\bibinfo {author} {\bibfnamefont {R.}~\bibnamefont
  {Foot}}\ and\ \bibinfo {author} {\bibfnamefont {X.~G.}\ \bibnamefont {He}},\
  }\href {https://doi.org/10.1016/0370-2693(91)90901-5} {\bibfield  {journal}
  {\bibinfo  {journal} {Phys. Lett. B}\ }\textbf {\bibinfo {volume} {267}},\
  \bibinfo {pages} {509} (\bibinfo {year} {1991})}\BibitemShut {NoStop}%
\bibitem [{\citenamefont {Holdom}(1986)}]{Holdom:1985ag}%
  \BibitemOpen
  \bibfield  {author} {\bibinfo {author} {\bibfnamefont {B.}~\bibnamefont
  {Holdom}},\ }\href {https://doi.org/10.1016/0370-2693(86)91377-8} {\bibfield
  {journal} {\bibinfo  {journal} {Phys. Lett. B}\ }\textbf {\bibinfo {volume}
  {166}},\ \bibinfo {pages} {196} (\bibinfo {year} {1986})}\BibitemShut
  {NoStop}%
\bibitem [{\citenamefont {Ellis}\ \emph {et~al.}(2018)\citenamefont {Ellis},
  \citenamefont {Fairbairn},\ and\ \citenamefont {Tunney}}]{Ellis:2018xal}%
  \BibitemOpen
  \bibfield  {author} {\bibinfo {author} {\bibfnamefont {J.}~\bibnamefont
  {Ellis}}, \bibinfo {author} {\bibfnamefont {M.}~\bibnamefont {Fairbairn}},\
  and\ \bibinfo {author} {\bibfnamefont {P.}~\bibnamefont {Tunney}},\ }\href
  {https://doi.org/10.1140/epjc/s10052-018-5738-6} {\bibfield  {journal}
  {\bibinfo  {journal} {Eur. Phys. J. C}\ }\textbf {\bibinfo {volume} {78}},\
  \bibinfo {pages} {238} (\bibinfo {year} {2018})},\ \Eprint
  {https://arxiv.org/abs/1705.03447} {arXiv:1705.03447 [hep-ph]} \BibitemShut
  {NoStop}%
\bibitem [{\citenamefont {Bauer}\ \emph {et~al.}(2017)\citenamefont {Bauer},
  \citenamefont {Neubert},\ and\ \citenamefont {Thamm}}]{Bauer:2018onh}%
  \BibitemOpen
  \bibfield  {author} {\bibinfo {author} {\bibfnamefont {M.}~\bibnamefont
  {Bauer}}, \bibinfo {author} {\bibfnamefont {M.}~\bibnamefont {Neubert}},\
  and\ \bibinfo {author} {\bibfnamefont {A.}~\bibnamefont {Thamm}},\ }\href
  {https://doi.org/10.1007/JHEP12(2017)044} {\bibfield  {journal} {\bibinfo
  {journal} {JHEP}\ }\textbf {\bibinfo {volume} {12}},\ \bibinfo {pages}
  {044}},\ \Eprint {https://arxiv.org/abs/1708.00443} {arXiv:1708.00443
  [hep-ph]} \BibitemShut {NoStop}%
\bibitem [{\citenamefont {Grzadkowski}\ \emph {et~al.}(2010)\citenamefont
  {Grzadkowski}, \citenamefont {Iskrzynski}, \citenamefont {Misiak},\ and\
  \citenamefont {Rosiek}}]{Grzadkowski:2010es}%
  \BibitemOpen
  \bibfield  {author} {\bibinfo {author} {\bibfnamefont {B.}~\bibnamefont
  {Grzadkowski}}, \bibinfo {author} {\bibfnamefont {M.}~\bibnamefont
  {Iskrzynski}}, \bibinfo {author} {\bibfnamefont {M.}~\bibnamefont {Misiak}},\
  and\ \bibinfo {author} {\bibfnamefont {J.}~\bibnamefont {Rosiek}},\ }\href
  {https://doi.org/10.1007/JHEP10(2010)085} {\bibfield  {journal} {\bibinfo
  {journal} {JHEP}\ }\textbf {\bibinfo {volume} {10}},\ \bibinfo {pages}
  {085}},\ \Eprint {https://arxiv.org/abs/1008.4884} {arXiv:1008.4884 [hep-ph]}
  \BibitemShut {NoStop}%
\bibitem [{\citenamefont {del Aguila}\ \emph {et~al.}(2010)\citenamefont {del
  Aguila}, \citenamefont {de~Blas},\ and\ \citenamefont
  {Perez-Victoria}}]{delAguila:2010mx}%
  \BibitemOpen
  \bibfield  {author} {\bibinfo {author} {\bibfnamefont {F.}~\bibnamefont {del
  Aguila}}, \bibinfo {author} {\bibfnamefont {J.}~\bibnamefont {de~Blas}},\
  and\ \bibinfo {author} {\bibfnamefont {M.}~\bibnamefont {Perez-Victoria}},\
  }\href {https://doi.org/10.1007/JHEP09(2010)033} {\bibfield  {journal}
  {\bibinfo  {journal} {JHEP}\ }\textbf {\bibinfo {volume} {09}},\ \bibinfo
  {pages} {033}},\ \Eprint {https://arxiv.org/abs/1005.3998} {arXiv:1005.3998
  [hep-ph]} \BibitemShut {NoStop}%
\bibitem [{\citenamefont {Efrati}\ \emph {et~al.}(2015)\citenamefont {Efrati},
  \citenamefont {Falkowski},\ and\ \citenamefont {Soreq}}]{Efrati:2015eaa}%
  \BibitemOpen
  \bibfield  {author} {\bibinfo {author} {\bibfnamefont {A.}~\bibnamefont
  {Efrati}}, \bibinfo {author} {\bibfnamefont {A.}~\bibnamefont {Falkowski}},\
  and\ \bibinfo {author} {\bibfnamefont {Y.}~\bibnamefont {Soreq}},\ }\href
  {https://doi.org/10.1007/JHEP07(2015)018} {\bibfield  {journal} {\bibinfo
  {journal} {JHEP}\ }\textbf {\bibinfo {volume} {07}},\ \bibinfo {pages}
  {018}},\ \Eprint {https://arxiv.org/abs/1503.07872} {arXiv:1503.07872
  [hep-ph]} \BibitemShut {NoStop}%
\bibitem [{\citenamefont {Falkowski}\ and\ \citenamefont
  {González-Alonso}(2021)}]{Falkowski:2019xoe}%
  \BibitemOpen
  \bibfield  {author} {\bibinfo {author} {\bibfnamefont {A.}~\bibnamefont
  {Falkowski}}\ and\ \bibinfo {author} {\bibfnamefont {M.}~\bibnamefont
  {González-Alonso}},\ }\href {https://doi.org/10.1007/JHEP04(2021)126}
  {\bibfield  {journal} {\bibinfo  {journal} {JHEP}\ }\textbf {\bibinfo
  {volume} {04}},\ \bibinfo {pages} {126}},\ \Eprint
  {https://arxiv.org/abs/1910.03002} {arXiv:1910.03002 [hep-ph]} \BibitemShut
  {NoStop}%
\bibitem [{\citenamefont {Jenkins}\ \emph {et~al.}(2013)\citenamefont
  {Jenkins}, \citenamefont {Manohar},\ and\ \citenamefont
  {Trott}}]{Jenkins:2013wua}%
  \BibitemOpen
  \bibfield  {author} {\bibinfo {author} {\bibfnamefont {E.~E.}\ \bibnamefont
  {Jenkins}}, \bibinfo {author} {\bibfnamefont {A.~V.}\ \bibnamefont
  {Manohar}},\ and\ \bibinfo {author} {\bibfnamefont {M.}~\bibnamefont
  {Trott}},\ }\href {https://doi.org/10.1007/JHEP10(2013)087} {\bibfield
  {journal} {\bibinfo  {journal} {JHEP}\ }\textbf {\bibinfo {volume} {10}},\
  \bibinfo {pages} {087}},\ \Eprint {https://arxiv.org/abs/1308.2627}
  {arXiv:1308.2627 [hep-ph]} \BibitemShut {NoStop}%
\bibitem [{\citenamefont {Jenkins}\ \emph {et~al.}(2014)\citenamefont
  {Jenkins}, \citenamefont {Manohar},\ and\ \citenamefont
  {Trott}}]{Jenkins:2013zja}%
  \BibitemOpen
  \bibfield  {author} {\bibinfo {author} {\bibfnamefont {E.~E.}\ \bibnamefont
  {Jenkins}}, \bibinfo {author} {\bibfnamefont {A.~V.}\ \bibnamefont
  {Manohar}},\ and\ \bibinfo {author} {\bibfnamefont {M.}~\bibnamefont
  {Trott}},\ }\href {https://doi.org/10.1007/JHEP01(2014)035} {\bibfield
  {journal} {\bibinfo  {journal} {JHEP}\ }\textbf {\bibinfo {volume} {01}},\
  \bibinfo {pages} {035}},\ \Eprint {https://arxiv.org/abs/1310.4838}
  {arXiv:1310.4838 [hep-ph]} \BibitemShut {NoStop}%
\bibitem [{\citenamefont {Alonso}\ \emph {et~al.}(2014)\citenamefont {Alonso},
  \citenamefont {Jenkins}, \citenamefont {Manohar},\ and\ \citenamefont
  {Trott}}]{Alonso:2013hga}%
  \BibitemOpen
  \bibfield  {author} {\bibinfo {author} {\bibfnamefont {R.}~\bibnamefont
  {Alonso}}, \bibinfo {author} {\bibfnamefont {E.~E.}\ \bibnamefont {Jenkins}},
  \bibinfo {author} {\bibfnamefont {A.~V.}\ \bibnamefont {Manohar}},\ and\
  \bibinfo {author} {\bibfnamefont {M.}~\bibnamefont {Trott}},\ }\href
  {https://doi.org/10.1007/JHEP04(2014)159} {\bibfield  {journal} {\bibinfo
  {journal} {JHEP}\ }\textbf {\bibinfo {volume} {04}},\ \bibinfo {pages}
  {159}},\ \Eprint {https://arxiv.org/abs/1312.2014} {arXiv:1312.2014 [hep-ph]}
  \BibitemShut {NoStop}%
\bibitem [{\citenamefont {Aebischer}\ \emph {et~al.}(2016)\citenamefont
  {Aebischer}, \citenamefont {Fael}, \citenamefont {Greub},\ and\ \citenamefont
  {Virto}}]{Aebischer:2015fzz}%
  \BibitemOpen
  \bibfield  {author} {\bibinfo {author} {\bibfnamefont {J.}~\bibnamefont
  {Aebischer}}, \bibinfo {author} {\bibfnamefont {M.}~\bibnamefont {Fael}},
  \bibinfo {author} {\bibfnamefont {C.}~\bibnamefont {Greub}},\ and\ \bibinfo
  {author} {\bibfnamefont {J.}~\bibnamefont {Virto}},\ }\href
  {https://doi.org/10.1007/JHEP05(2016)037} {\bibfield  {journal} {\bibinfo
  {journal} {JHEP}\ }\textbf {\bibinfo {volume} {05}},\ \bibinfo {pages}
  {037}},\ \Eprint {https://arxiv.org/abs/1512.02830} {arXiv:1512.02830
  [hep-ph]} \BibitemShut {NoStop}%
\bibitem [{\citenamefont {Aebischer}\ \emph {et~al.}(2017)\citenamefont
  {Aebischer}, \citenamefont {Gonzalez-Alonso}, \citenamefont {Jenkins},
  \citenamefont {Manohar},\ and\ \citenamefont {Stoffer}}]{Aebischer:2017ugx}%
  \BibitemOpen
  \bibfield  {author} {\bibinfo {author} {\bibfnamefont {J.}~\bibnamefont
  {Aebischer}}, \bibinfo {author} {\bibfnamefont {M.}~\bibnamefont
  {Gonzalez-Alonso}}, \bibinfo {author} {\bibfnamefont {E.~E.}\ \bibnamefont
  {Jenkins}}, \bibinfo {author} {\bibfnamefont {A.~V.}\ \bibnamefont
  {Manohar}},\ and\ \bibinfo {author} {\bibfnamefont {P.}~\bibnamefont
  {Stoffer}},\ }\href {https://doi.org/10.1007/JHEP09(2017)123} {\bibfield
  {journal} {\bibinfo  {journal} {JHEP}\ }\textbf {\bibinfo {volume} {09}},\
  \bibinfo {pages} {123}},\ \Eprint {https://arxiv.org/abs/1706.03783}
  {arXiv:1706.03783 [hep-ph]} \BibitemShut {NoStop}%
\bibitem [{\citenamefont {Abbaneo}\ \emph {et~al.}(2003)\citenamefont {Abbaneo}
  \emph {et~al.}}]{LEP:2003aa}%
  \BibitemOpen
  \bibfield  {author} {\bibinfo {author} {\bibfnamefont {D.}~\bibnamefont
  {Abbaneo}} \emph {et~al.} (\bibinfo {collaboration} {CERN LEP
  Collaborations}),\ }\href@noop {} {\bibfield  {journal} {\bibinfo  {journal}
  {CERN-EP/2003-091}\ } (\bibinfo {year} {2003})},\ \bibinfo {note} {uRL:
  https://cds.cern.ch/record/690329}\BibitemShut {NoStop}%
\bibitem [{\citenamefont {Altmannshofer}\ \emph {et~al.}(2014)\citenamefont
  {Altmannshofer}, \citenamefont {Gori}, \citenamefont {Pospelov},\ and\
  \citenamefont {Yavin}}]{Altmannshofer:2014pba}%
  \BibitemOpen
  \bibfield  {author} {\bibinfo {author} {\bibfnamefont {W.}~\bibnamefont
  {Altmannshofer}}, \bibinfo {author} {\bibfnamefont {S.}~\bibnamefont {Gori}},
  \bibinfo {author} {\bibfnamefont {M.}~\bibnamefont {Pospelov}},\ and\
  \bibinfo {author} {\bibfnamefont {I.}~\bibnamefont {Yavin}},\ }\href
  {https://doi.org/10.1103/PhysRevLett.113.091801} {\bibfield  {journal}
  {\bibinfo  {journal} {Phys. Rev. Lett.}\ }\textbf {\bibinfo {volume} {113}},\
  \bibinfo {pages} {091801} (\bibinfo {year} {2014})},\ \Eprint
  {https://arxiv.org/abs/1406.2332} {arXiv:1406.2332 [hep-ph]} \BibitemShut
  {NoStop}%
\bibitem [{\citenamefont {Altmannshofer}\ \emph {et~al.}(2019)\citenamefont
  {Altmannshofer}, \citenamefont {Mart{\'i}nez},\ and\ \citenamefont
  {Pospelov}}]{Altmannshofer:2019zhy}%
  \BibitemOpen
  \bibfield  {author} {\bibinfo {author} {\bibfnamefont {W.}~\bibnamefont
  {Altmannshofer}}, \bibinfo {author} {\bibfnamefont {D.}~\bibnamefont
  {Mart{\'i}nez}},\ and\ \bibinfo {author} {\bibfnamefont {M.}~\bibnamefont
  {Pospelov}},\ }\href {https://doi.org/10.1103/PhysRevD.100.115029} {\bibfield
   {journal} {\bibinfo  {journal} {Phys. Rev. D}\ }\textbf {\bibinfo {volume}
  {100}},\ \bibinfo {pages} {115029} (\bibinfo {year} {2019})},\ \Eprint
  {https://arxiv.org/abs/1909.02089} {arXiv:1909.02089 [hep-ph]} \BibitemShut
  {NoStop}%
\bibitem [{\citenamefont {Anthony}\ \emph {et~al.}(2005)\citenamefont {Anthony}
  \emph {et~al.}}]{Anthony:2005pm}%
  \BibitemOpen
  \bibfield  {author} {\bibinfo {author} {\bibfnamefont {P.~L.}\ \bibnamefont
  {Anthony}} \emph {et~al.} (\bibinfo {collaboration} {SLAC E158}),\ }\href
  {https://doi.org/10.1103/PhysRevLett.95.081601} {\bibfield  {journal}
  {\bibinfo  {journal} {Phys. Rev. Lett.}\ }\textbf {\bibinfo {volume} {95}},\
  \bibinfo {pages} {081601} (\bibinfo {year} {2005})},\ \Eprint
  {https://arxiv.org/abs/hep-ex/0504049} {hep-ex/0504049} \BibitemShut
  {NoStop}%
\bibitem [{\citenamefont {Benesch}\ \emph {et~al.}(2014)\citenamefont {Benesch}
  \emph {et~al.}}]{Benesch:2014bas}%
  \BibitemOpen
  \bibfield  {author} {\bibinfo {author} {\bibfnamefont {J.}~\bibnamefont
  {Benesch}} \emph {et~al.} (\bibinfo {collaboration} {MOLLER Collaboration}),\
  }\href@noop {} {\  (\bibinfo {year} {2014})},\ \Eprint
  {https://arxiv.org/abs/1411.4088} {arXiv:1411.4088 [nucl-ex]} \BibitemShut
  {NoStop}%
\bibitem [{\citenamefont {Jegerlehner}\ and\ \citenamefont
  {Nyffeler}(2009)}]{Jegerlehner:2009ry}%
  \BibitemOpen
  \bibfield  {author} {\bibinfo {author} {\bibfnamefont {F.}~\bibnamefont
  {Jegerlehner}}\ and\ \bibinfo {author} {\bibfnamefont {A.}~\bibnamefont
  {Nyffeler}},\ }\href {https://doi.org/10.1016/j.physrep.2009.04.003}
  {\bibfield  {journal} {\bibinfo  {journal} {Phys. Rept.}\ }\textbf {\bibinfo
  {volume} {477}},\ \bibinfo {pages} {1} (\bibinfo {year} {2009})},\ \Eprint
  {https://arxiv.org/abs/0902.3360} {arXiv:0902.3360 [hep-ph]} \BibitemShut
  {NoStop}%
\bibitem [{\citenamefont {Endo}\ \emph {et~al.}(2021)\citenamefont {Endo},
  \citenamefont {Mishima},\ and\ \citenamefont {Ueda}}]{Endo:2021zal}%
  \BibitemOpen
  \bibfield  {author} {\bibinfo {author} {\bibfnamefont {M.}~\bibnamefont
  {Endo}}, \bibinfo {author} {\bibfnamefont {G.}~\bibnamefont {Mishima}},\ and\
  \bibinfo {author} {\bibfnamefont {T.}~\bibnamefont {Ueda}},\ }\href
  {https://doi.org/10.1007/JHEP10(2021)179} {\bibfield  {journal} {\bibinfo
  {journal} {JHEP}\ }\textbf {\bibinfo {volume} {10}},\ \bibinfo {pages}
  {179}},\ \Eprint {https://arxiv.org/abs/2105.10445} {arXiv:2105.10445
  [hep-ph]} \BibitemShut {NoStop}%
\end{thebibliography}%

\end{document}